\documentclass[hyperref,solaromanenum,lines,longbibliography]{spr-sola}

\usepackage{ifluatex}
\ifluatex
\usepackage{pdftexcmds}
\makeatletter
\let\pdfstrcmp\pdf@strcmp
\let\pdffilemoddate\pdf@filemoddate
\fi
\usepackage{comment}
\usepackage{soul}
\usepackage{hyperref}

\usepackage{graphicx}                    
\usepackage{color}                       

\chardef\us=`\_
\graphicspath{{./}{figures/}}

\begin{document}

\begin{frontmatter}

\title{Detection of ``diffuse'' coronal \ion{He}{1} 1083 during the April 8 2024 Solar Eclipse: evidence for terrestrial atmospheric scattering contribution}

\author[addressref={aff1,aff2},email={momchil.molnar@swri.org}]{\inits{M.}\fnm{Momchil E.}~\snm{Molnar}\orcid{0000-0003-0583-0516}}
\author[addressref={aff1},email={casini@ucar.edu}]{\inits{R.}\fnm{Roberto}~\snm{Casini}\orcid{0000-0001-6990-513X}}
\author[addressref={aff1},email={pbryans@ucar.edu}]{\inits{P.}\fnm{Paul}~\snm{Bryans}\orcid{0000-0001-5681-9689}}
\author[addressref={aff1},email={berkey@ucar.edu}]{\inits{B.}\fnm{Ben}~\snm{Berkey}\orcid{0000-0000-0000-0000}}
\author[addressref={aff1},email={kalistatyson@gmail.com}]{\inits{K.}\fnm{Kalista}~\snm{Tyson}\orcid{0000-0000-0000-0000}}

\address[id=aff1]{High Altitude Observatory, 
	NSF National Center for Atmospheric Research, Boulder, CO, USA}
\address[id=aff2]{Southwest Research Institute, Boulder, CO, USA}



\begin{abstract}

Strong \ion{He}{1} 1083\,nm atomic line signals have previously been measured 
during total solar eclipses at coronal heights above the lunar limb. These
rather unexpected measurements have kindled a discussion about 
the suggested presence of significant amounts of neutral helium at coronal conditions.
We performed spectroscopic observations of the \ion{He}{1} 1083\,nm wavelength region
during the April 8th 2024 total
solar eclipse, using an instrument specifically designed to test the presence of \ion{He}{1} 1083 in the solar corona.
We were able to detect the \ion{He}{1} 1083 line, the forbidden coronal line of \ion{Fe}{13} at 1074.7\,nm, as well as the
chromospheric \ion{H}{1} 1093.8\,nm (Paschen-$\gamma$) line in our data.
The chromospheric \ion{He}{1} 1083\,nm and \ion{H}{1} 1093.8\,nm
lines were detected in both the corona and on the lunar disc.
We hypothesize that our observations support a terrestrial atmospheric scattering of the solar flash spectrum as the  origin of the 
\ion{He}{1} 1083 signal during the April 8th 2024 eclipse. Our findings 
challenge the notion of abundant neutral helium in the solar corona 
suggested by previous eclipse observations.

\end{abstract}

\end{frontmatter}



\section{Introduction} 
\label{sec:intro}

Past eclipse and coronagraphic observations have shown the presence of 
\ion{He}{1} 1083.0\,nm (\ion{He}{1} 1083) atomic line emission
in the solar corona, 
far away from any notable prominence contribution \citep{1996ApJ...456L..67K}. In particular, the eclipse 
observations from
\citet{1996ApJ...456L..67K} show an omnipresent \ion{He}{1} 1083
signal up to a few solar radii above the limb, with the line width corresponding to 
coronal temperatures, as was confirmed in following eclipse expeditions \citep{Dima_2016}. This is a rather surprising finding, given that most of the helium should be fully ionized under coronal conditions
\citep{2020ApJ...898...72D}.
However, coronagraphic observations with the Solar-C 
telescope at the Haleakala High Altitude Observatory, Hawaii  \citep{2007ApJ...667L.203K,2010ApJ...722.1411M} detected
\ion{He}{1} 1083 signal during Solar maximum in isolated coronal regions.
More recently, \citet{2019ApJ...877...10J} detected ubiquitous \ion{He}{1} 1083 emission during
the Great American Solar eclipse of August 21 2017.
The origin of the \ion{He}{1} has been widely
debated in the literature, given that the presence of a significant amount of neutral 
helium in the millions degree corona is physically puzzling, while the limited ground-based detections suffer from systematic uncertainties
\citep{2010ApJ...722.1411M}. 

More recently, the METIS coronograph on Solar Orbiter
\citep{SolO_2020, METIS_2020} observed prominence eruption material at 
distances greater than 5\,R$_{\odot}$, detecting polarized
signals potentially originating from
the \ion{He}{1} D$_3$ 587.6\,nm line \citep{2023ApJ...957L..10H} in the prominence
material, but not everywhere in the corona. These METIS observations, together with other eclipse observations such as the ones presented by \cite{2017ApJ...842L...7D}, show the localized nature of the \ion{He}{1} emission to prominence material in the solar corona, taken from outside of eclipse measurements. 
The lower energy
state of the \ion{He}{1} D$_3$ transition is the same as the  upper state of the \ion{He}{1} 1083 line
\citep[e.g. Figure 4 in][]{1982SoPh...79..291L}. From the modeling perspective, recent
numerical work by \citet{2020ApJ...898...72D} presents a compelling evidence for
bright \ion{He}{1} 1083 signals in the lower corona similar in brightness to the 
routinely observed \ion{Fe}{13} 1074/1079\,nm line pair \citep{2008SoPh..247..411T}. 
This work showed how calculations including dielectronic recombination allow
for a significant population of neutral helium in the lower corona, but
does not explain the extensive (up to a few solar radii) \ion{He}{1} 1083 
emission observed in \citet{1996ApJ...456L..67K}. 
The aforementioned evidence for coronal \ion{He}{1} 1083 presents an
intriguing puzzle for what physical mechanism may be responsible for such an emission, 
and in particular if this phenomenon is of solar origin. 

Detecting coronal \ion{He}{1} 1083 is an exciting prospect
for coronal magnetometry, based on the sensitivity of the \ion{He}{1} 1083
to the coronal magnetic field through the
Hanle effect \citep{2016FrASS...3...20R}. Most of the previous 
coronal Hanle magnetometry 
work has been on the Hanle effect in the ultraviolet (UV) region of 
the solar spectrum \citep{1999A&A...345..999R,2002A&A...396.1019R}. 
Many permitted resonance lines in the UV are 
Hanle sensitive to magnetic fields slightly higher than the 
ones found in the solar corona \citep{2022SoPh..297...96K,2024ApJ...971...27K}. 
However, UV polarimetry is not available for routine solar 
observations due to the requirement for space-based observatory with
challenging polarimetric design
\citep{2023BAAS...55c.047C,2023BAAS...55c.048C}.
The advantage of using \ion{He}{1} 1083 for Hanle-based 
coronal magnetometry is the combination of its brightness and location in the 
infrared (IR) part of the spectrum, making it readily observable with current 
instrumentation \citep{Centeno_2008, Molnar_2024}. 
Furthermore, the \ion{He}{1} 1083 magnetic field sensitivity regime is very
similar to the expected magnetic field strengths in the quiet-Sun corona 
\citep{2020Sci...369..694Y} and in erupting solar prominences at 
coronal heights~\citep{2018ApJ...862...54F}. 

However, there is the well established observational finding
that strong chromospheric
lines (such as the \ion{Ca}{2} H/K and the \ion{H}{1} Balmer series)
are observed in the solar corona during total solar eclipses in the blue part 
of the visible spectrum \citep[e.g. ][]{Grotrian_1934,Colacevic_1956,1955C&T....71..288M,1964ApJ...140..313D} due to
non-solar origins.
This puzzling finding dates back to the beginning of the 20th century 
and invited two contrasting physical explanations:  
emission from cool, ``diffuse'', prominence material dispersed in the million degree corona
\citep{1972SoPh...26..366B};
or scattering of chromospheric flash spectrum in the local environment around Earth, redirecting chromospheric line photons in the eclipse umbra \citep{1971SoPh...17...89C,1972SoPh...26..366B}.
An eclipse experiment by \citet{Stellmacher_1974} 
showed strong evidence for the 
non-coronal origin of the observed chromospheric lines in the blue
part of the spectrum, by simultaneously detecting chromospheric line
signal both on the lunar
disc and in the solar corona during totality. \citet{Stellmacher_1974} demonstrated that
the cool chromospheric lines in the blue part of the spectrum, seen on the lunar disc, have a line width 
dependence with their rest wavelength that follows a relation expected for signals 
originated from Rayleigh scattering in the terrestrial atmosphere. The proposed explanation is that the flash spectrum, seen by the atmosphere close to the observer (see Figure 4 of \citet{Stellmacher_1974}) scatters twice in the atmosphere and is detected as a coronal signal by the observers instrument which resides in the eclipse shadow. The 
recent eclipse observations by \citet{2019ApJ...877...10J} show a brightening of
the \ion{He}{1} 1083 emission in the corona 
closer to second and third contacts, which
agrees with the hypothesis of a scattering origin in the Earth atmosphere of the flash spectrum. 
With the recent growth of eclipse spectroscopy 
\citep{2019A&A...632A..86K, 2023SoPh..298...75M, 2024ApJ...966..122Z} 
understanding the atmospheric effects on the spectral information detected during totality becomes of utmost importance.

The yet unsettled state of knowledge about 
the origin of the ``diffuse'' 
coronal \ion{He}{1} 1083 presented above
motivated us to perform eclipse measurements of the near-IR coronal spectrum
during the April 8 2024 eclipse to test the origin of the \ion{He}{1} 1083
eclipse signals. To this aim, we designed and built the Coronal HElium 
Emission Spectrograph Experiment (CHEESE), a spectrograph with the necessary spectral 
resolution and field-of-view necessary to resolve the line width of \ion{He}{1} 
\citep{1996ApJ...456L..67K} to test the hypothesis of \ion{He}{1} 1083 origin from 
atmospheric scattering of the flash spectrum.
In this paper we present the results from CHEESE observations taken during the NCAR/HAO April 8 2024 eclipse expedition. We describe both the
instrument design and the observational campaign in Section~\ref{sec:Observations}.
We present our interpretation of the results and our findings in Section~\ref{sec:Results} and their implications in
Section~\ref{sec:Conclusions}.

\section{Observations} 
\label{sec:Observations}

\subsection{Instrument design}
\label{sec:instrument_design}

\begin{figure}[ht!]
	\includegraphics[width=\textwidth]{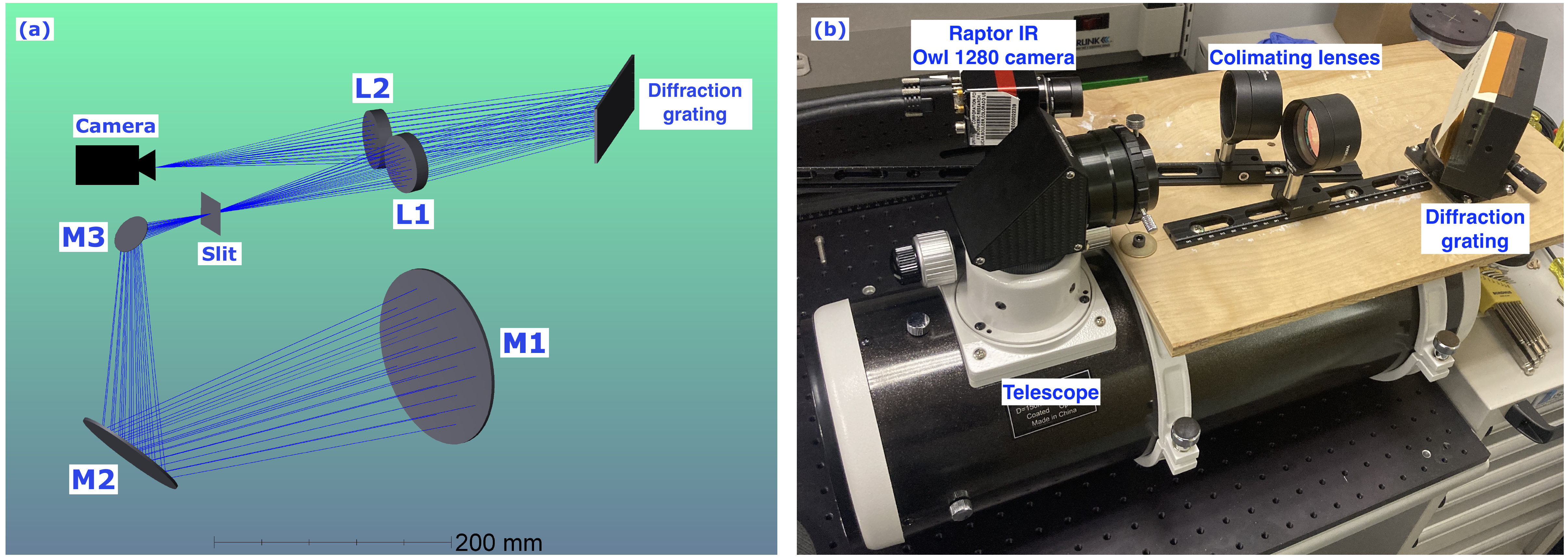}
	\caption{Overview of the CHEESE instrument. \textbf{Panel (a)}: Optical layout of the CHEESE instrument produced with the Zeemax
    software suite. The primary and secondary
		mirrors of the Skywatcher Quattro telescope 
		correspond to \emph{M1} and \emph{M2}, 
        the diagonal 90$^{\circ}$ mirror
        to \emph{M3}, and the two collimating lenses in 
        the spectrograph are labeled as \emph{L1} and \emph{L2}.
     \textbf{Panel (b)}: CHEESE during laboratory bench assembly in NSF NCAR/HAO. The essential optical
 		components and the camera are labeled. Note that during the actual observations the
 		spectrograph was housed inside a thermally insulated light-tight enclosure 
 		to isolate it from the ambient environment.}
	\label{fig:CHEESE_optical_layout}
\end{figure}

CHEESE is a diffraction grating spectrograph 
operating in the near-IR designed to reproduce the previous observations 
of \ion{He}{1} 1083 in the corona
\citep{1996ApJ...456L..67K}. The optical layout of the system is presented in
Figure~\ref{fig:CHEESE_optical_layout} panel (a). An actual photo of
the assembled spectrograph is shown 
in Figure~\ref{fig:CHEESE_optical_layout} panel (b), with some of the 
essential components labeled. We used a commercially available f/4
Newtonian Skywatcher Quattro 150P telescope with primary mirror of D=15\,cm
and focal length of 600\,mm. A 40 $\mu$m slit was placed in the primary focus of
the telescope while mounted on a helical focuser for fine position adjustments.
We used a Bausch and Lomb diffraction grating, with a blaze angle
of 17.5$^{\circ}$ in inverse configuration (i.e., at 72.5$^{\circ}$ blaze)
with 600 lines/mm. The grating was mounted on a 3D-printed adapter to a
rotational stage for fine adjustment of the beam incidence angle
(and the resultant observed wavelength range).
We use two near-IR ThorLabs achromatic lenses (L1 and L2 in 
Figure~\ref{fig:CHEESE_optical_layout}) with 200 
mm focal length to illuminate the grating with the
collimated beam and then image the dispersed spectrum on the camera. 
\texttt{Raptor IR Owl 1280} InGaS camera was used for this experiment, suitable with its high quantum efficiency (QE\,$>$\,90\% at 1\,$\mu$m), large 10\,$\mu$m 
pixel size, and low readout and thermal noise properties.
The camera has active internal cooling, which we enhanced with externally 
mounted radiators. This allowed for the stable operation of the camera 
with the detector cooled to a 1$^{\circ}$\,C temperature. Despite the ability 
of the camera to sustain lower temperatures, we decided against running it below
freezing to avoid any condensation buildup on its entrance window.
 
We operated CHEESE in the second diffraction order for the best combination of 
efficiency and dispersion. To reject overlapping orders on the detector, an 
additional 980 nm high pass order sorting filter is required, which was
mounted right before the camera body, to remove 
contribution from higher orders; on the upper end of the
wavelength range, the camera QE efficiency drops to effectively zero above 1.8 $\mu$m, 
which rejects the overlapping lower order contributions. The resulting
spectrograph has predicted spectral resolution of \textbf{R}\,$\sim$\,9,600 and
a plate scale of 3.43\,{\arcsec}/pixel.
Given an estimated vignetting of 0.9, we estimated \ion{He}{1} 1083 
signal-to-noise ration (SNR) of a few hundred after 1 minute of 
integration under coronal conditions, such as the ones described in \citet{1996ApJ...456L..67K}. The 1\,cm slit of 
CHEESE spans 58\,{\arcmin} on the plane of the sky, allowing for sampling of 
both the lunar disc and the corona simultaneously, as shown in 
Figure~\ref{fig:AIA_overview}.
CHEESE was mounted on a \texttt{Losmandy GM-8} equatorial mount with a
\texttt{Gemini-1} guiding system; the telescope and spectrograph 
weigh about 23 pounds, well below the limit of the 
guiding system and light enough 
for swift (re-)deployment, as proved necessary during our
expedition. The 3D printed slit and diffraction grating
mount designs are publicly available in the repository associated with this
publication.\footnote{\url{https://github.com/momomolnar/CHEESE-data}}

\subsection{Observational campaign}

\begin{figure}[ht!]
	\includegraphics[width=\textwidth]{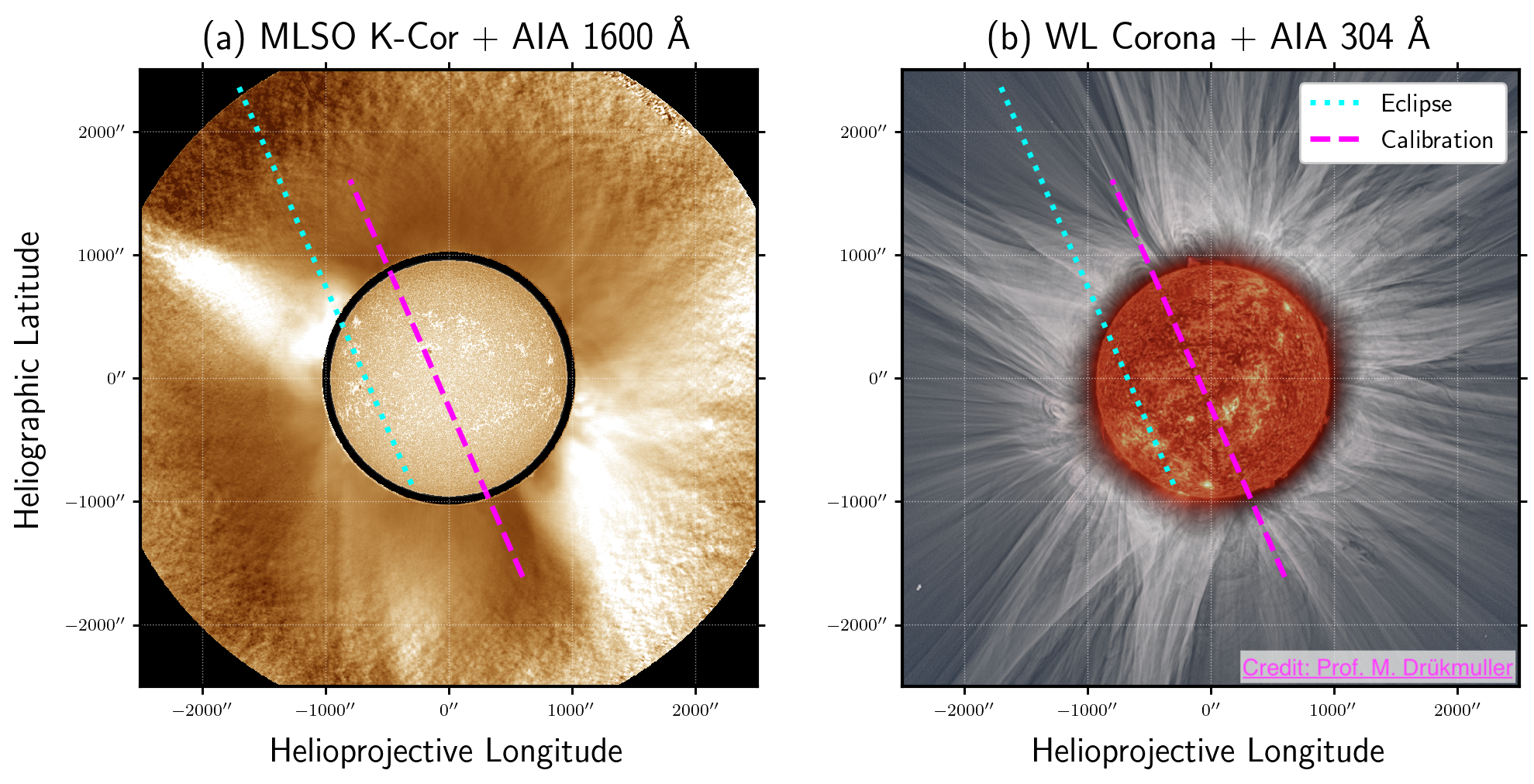}
	\caption{Composite context coronal images showing the position of the CHEESE slit during the eclipse campaign:
    \textbf{Panel\,(a):}
            solar disc data taken from SDO/AIA 1600\,{\AA} channel;
            the corona above 1.01\,R$_{\odot}$ is imaged by 
            the MLSO/K-Cor white light coronograph on April 9 2024, the day after the eclipse.
    \textbf{Panel\,(b)}:
            Solar disc taken from SDO/AIA 304\,{\AA} channel;
            the corona above is imaged in white light from Durango, Mexico; the image is created by Prof. Miloslav Druckm\"{u}ller and is reproduced with his permission.
            The magenta dashed line 
            corresponds to the approximate slit location during the
            CHEESE calibration and the cyan dotted line shows the approximate 
            slit location during totality.
            }
	\label{fig:AIA_overview}
\end{figure}

\begin{figure}
\includegraphics[width=\textwidth]{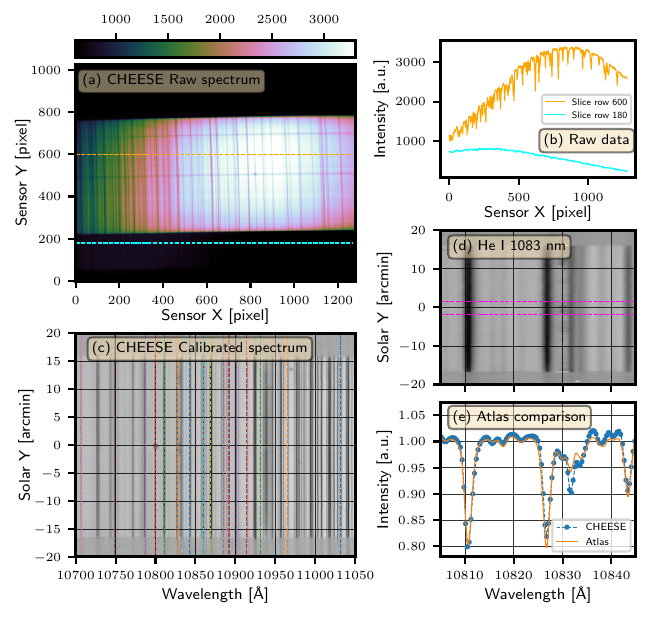}
\caption{Calibration on-disc data taken at 13:23 UT Apr 8 2024. \textbf{Panel (a):} Raw CHEESE spectrum in sensor coordinates. 
The X-direction corresponds roughly to the dispersion direction and the Y-axis to the slit direction. \textbf{Panel (b):} Slices through the spectrum of the disc (orange) and the sky (blue) (corresponding to the dashed lines in Panel (a)).
    \textbf{Panel (c)}: Flat fielded, continuum normalized, and de-rotated CHEESE spectrum 
	with dispersion correction applied, where the intensity of each row (Solar-Y) is normalized to unity. The absorption lines used for the dispersion and
	rotation corrections are marked as thin dashed lines. \textbf{Panel (d):} 
	the region around the \ion{He}{1} 1083, where the \ion{Si}{1} 10827\,{\AA}
	and \ion{He}{1} 1083 lines are clearly seen. 
    The dashed magenta lines mark the region used for 
    studying the average spectral properties in the next panel. 
    \textbf{Panel (e):} Averaged CHEESE observation (blue dots) compared with an instrument model based on 
    solar atlas data (dashed orange).}
\label{fig:calibration_data}
\end{figure}

The NCAR/HAO expedition observed the total solar eclipse 
from Dardanelle, Arkansas, USA
at geographical coordinates N 35$^{\circ}$12{\arcmin}14.7$\arcsec$ 
W 93$^{\circ}$12{\arcmin}16.5{\arcsec}. Our original intent was to 
observe the eclipse from a location close to San Antonio, Texas, based on
the historically favorable climatological record. However, adverse weather forecasts 
led us to relocate our eclipse observing campaign to Dardanelle, AR. 
We set our instruments on the night before the eclipse (April 7-8 2024) with 
clear skies allowing us to align our mounts. The weather during the eclipse was 
stable with thin high cloud cover present during totality. 

Overview of the solar coronal conditions on the day of the eclipse is shown in
Figure~\ref{fig:AIA_overview}. The composite images consist of photospheric and chromospheric  data from SDO/AIA 
\citep{2012SoPh..275...17L} from April 8 2024 
and the whitelight coronograph/ white light images are from: Panel (a) the Mauna Loa Solar Observatory 
(MLSO) K-Cor~\citep{deWijn_2012} from April 9 2024; Panel (b): white light coronal image taken during totality taken by Prof. Miloslav Drukmuller in Durango, Mexico. The latter image is reproduced with the permission of the author and can be found on his \href{http://www.zam.fme.vutbr.cz/~druck/Eclipse/Ecl2024u/Durango/TSE_2024_400mm_304A/0-info.htm}{personal webpage}. The
MLSO K-Cor observes the whitelight continuum corona in the region of 780\,nm
and the data product which we show in Figure~\ref{fig:AIA_overview} has 
a normalized radial gradient filter applied. 
We have indicated on Figure~\ref{fig:AIA_overview} the approximate locations 
of the CHEESE slit during the calibration process as the dashed magenta line,
and its approximate location during the eclipse observations as the dotted cyan
one. 

The CHEESE data were calibrated by initially removing the dark frames and
then dividing out a flat field. A preview of a raw frame from CHEESE is shown in Figure~\ref{fig:calibration_data} Panel (a). The sunspot group associated 
with NOAA AR 13628, located close to disc center at approximately Y-pixel coordinate 
500, was used as a fiducial line to remove the rotation of the 
spectrum on the detector,
which was measured to be about 1.25\,$^{\circ}$. The wavelength dispersion of 
the spectrograph was then determined by finding the precise location of 17 known 
spectral infrared lines using the atlas 
of \cite{Dellbouille_IR_solar_atlas}\footnote{The atlas is hosted on the 
BASS2000 webpage:\,\href{https://bass2000.obspm.fr/}{https://bass2000.obspm.fr/}}. 
We found an anomalous vertical column shift (along the slit) on the left 
side of the detector, which was removed after finding the center of the slit image across the detector.
The spectral lines used for calibration
are shown in panel (c) of Figure~\ref{fig:calibration_data} as the thin
dashed colored lines. We chose the particular calibration 
lines to be well separated from surrounding (complex) spectral features 
and blends. We computed the dispersion of our spectrograph by 
fitting the centroids of the reference spectral lines and then
fitting a parabolic dispersion relationship for each row of the detector.
Then, for every image row the wavelength 
dependence was used to resample the data uniformly in wavelength. 
The resulting uniformly resampled spectrum is shown in panel (c) of Figure~\ref{fig:calibration_data} where the continuum of each row has been normalized to unity. The wavelength dispersion
mapping we found was smooth across the detector, showing the fidelity of the 
recovered spectra. Radiometric calibration was attempted with an 
IR-calibrated opal from the NCAR/HAO collection, but we do not use 
those data in this work due to camera malfunction during obtaining those data.

A magnified part of the solar disc spectrum centered
on the \ion{He}{1} 1083 region is shown in panel (d) of 
Figure~\ref{fig:calibration_data}. It demonstrates
the presence (and detectability) of \ion{He}{1} 1083 as well as the nearby
photospheric \ion{Si}{1} 1082.7\,nm lines observed on the 
solar disc. In particular, note the \ion{He}{1} 1083 line is in emission above 
the solar limb, noted as the brightening in it 
in the continuum averaged spectrum in panel (d) of Figure~\ref{fig:calibration_data}. 
In panel (e) of Figure~\ref{fig:calibration_data} we present a comparison between the CHEESE
data and the solar atlas from \citet{1999SoPh..184..421N}.
The average spectrum from the CHEESE data between the 
magenta curves in panel 
Figure~\ref{fig:calibration_data} (d) is shown as the blue dots in panel (e),
and the solar atlas fits to those data are presented as the orange line. We fitted 
the atlas to the CHEESE data by rescaling the atlas
intensity level, including a scattered light component (flat background),
and degrading it with a Gaussian line spread function with its width 
kept as a free parameter.
The resulting fit provides an estimate of the effective resolving power 
of the spectrograph as well as an estimate of the amount of scattered light, 
based on the flat (gray) component needed to be included to fit the spectrum 
to the atlas. 
Based on the fitting of the solar atlas spectrum to our measurements we 
estimated that CHEESE was observing with an effective spectral resolution
of about R\,$\sim$\,7000. This effective spectral resolution shows that our 
spectrograph behaves similarly to its designed specification; 
most importantly it is able to detect (and resolve) the 
\ion{He}{1} 1083 line successfully. 

\section{Results from the eclipse observations} 
\label{sec:Results}

\begin{figure}[ht!]
	\includegraphics[width=1.0\textwidth]{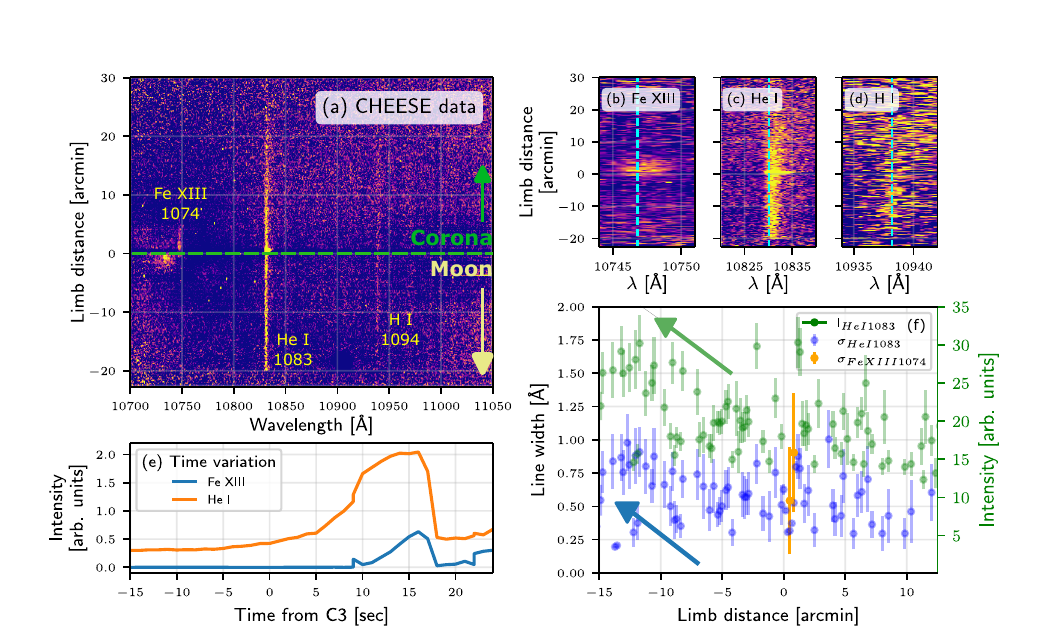}
	\caption{Results from the eclipse observations. \textbf{Panel (a)}: Processed CHEESE spectrum obtained during the 
	total solar eclipse between 11:54:17 - 11:54:28 UT on April 8th 2024. In total we had. Note
	the ever present \ion{He}{1} 1083.0\,nm line, as well as the 
	\ion{Fe}{13} 1074\,nm line. A very dim spectral line is detected at the precise location of the chromospheric \ion{H}{1} 
	Paschen-$\gamma$ 1093.8\,nm line. The dashed green line marks the location of the lunar limb. \textbf{Panels (b)-(d)}: 
	Zoom in over the detected spectral lines in the CHEESE data, in order: (b) \ion{Fe}{13} 1074.7\,nm; 
	(c) \ion{He}{1} 1083.0\,nm; (d) \ion{H}{1} Paschen-$\gamma$ 1093.8\,nm. The cyan line marks the 
	reference wavelength. \textbf{Panel (e)}: Time variation of the \ion{He}{1} 1083 and \ion{Fe}{13} 1074 line intensity around 3rd contact (C3); \textbf{Panel (f)}: Line width in Angstrom of the detected  \ion{He}{1} 1083 
	(in blue) and \ion{Fe}{13} 1074\,nm (in orange), and the \ion{He}{1} 1083 line intensity (in green).} 
	\label{fig:results_spectra}
\end{figure}

During the eclipse, we pointed our instrument towards the east solar limb
where a bright streamer 
was present on the day of the eclipse, as shown in Figure~\ref{fig:AIA_overview}. 
We chose this pointing to maximize the likelihood of detecting
coronal \ion{He}{1} signal, by observing a coronal region of enhanced
plasma density.

The CHEESE data collected during the eclipse totality suffered from 
a dynamic range 
issue due to the internal data reduction performed onboard the camera. 
This step reduced the camera dynamic range to be effectively 12-bit 
compared to the 16-bit detector well depth.
Since our slit was crossing the chromosphere,
bright chromospheric material ($\sim$10$^5$\,$\mu$B$_{\odot}$) in 
the \ion{He}{1} 1083 drove the response of 
the camera such that the anticipated weak coronal signals 
($\sim$10\,$\mu$B$_{\odot}$) were outside of the effective dynamic range 
of the camera. Even so, CHEESE managed to detect the \ion{Fe}{13} 1074 nm line during the eclipse with very low signal-to-noise ratio (SNR) close to the lunar limb.
During the time period around third contact, the Moon covered almost all of 
the previously visible bright chromospheric signal and our measurements 
started to exhibit clearly the 
coronal \ion{Fe}{13} 1074\,nm line, as well as the other spectral 
signals described below. We co-added 125 exposures (10 seconds of effective 
exposure time) taken between 18:54:17-18:54:28, which is 10 seconds after third 
contact for the exact observing location. Since we were observing on the east limb 
during third contact, our slit was not crossing the flash spectrum.

 The CHEESE data from the totality are shown in Figure~\ref{fig:results_spectra}. We have applied the exact same reduction
 steps as for the data shown in Figure~\ref{fig:CHEESE_optical_layout}, where we have 
 used the wavelength dispersion relation from our on disc observations. Panel (a) 
of Figure~\ref{fig:results_spectra} shows the full detector spectrum averaged
over the time period between 18:54:17-18:54:28 UT. 
The approximate green line shows the approximate position of the lunar limb,
where one can clearly see the \ion{He}{1} 1083 line, observed both in the 
corona and on the lunar disk. However, the \ion{Fe}{13} 1074\,nm line
is detected only in the coronal region detected above the lunar limb. The temporal variation line intensity variation of the \ion{He}{1} 1083 and 
the \ion{Fe}{13} 1074 lines is displayed in  Figure~\ref{fig:results_spectra} panel (e) around third contact (C3). We have computed the line intensity as the integrated signal from the coronal region of the detector where the two lines are observed, with an estimate for the continuum subtracted from equal 
area windows, located close to the nominal spectral line location. We have also applied 10 frame temporal averaging, resulting in effective 
temporal resolution of 0.8\,s. Note that 
after 3rd contact, we can detect both lines well, until the sky brightness increases significantly and both lines are not distinguishable from the sky background, at about 18 seconds after C3.
Note that over the temporal averaging window, the lunar limb moves about 
1.2\,{\arcmin};
hence care should be taken in the interpretation of the region close to the 
limb. Since we do not have a slitjaw camera to provide context imaging, we were not able to align the consecutive frames, but the aforementioned phenomenon should not affect the coronal data farther from the limb. 
The \ion{H}{1} 1093.8\,nm Paschen-$\gamma$ line is detected clearly on the lunar 
 disc and not so much in the coronal region, due to the increased background and noise 
 levels, as shown in panels 
 (a) and (d) in the Figure~\ref{fig:results_spectra}. We note that the 
 detection of those lines is further supported by the precise coincidence
 of their detected wavelength locations, as shown in panels (b)-(d) of 
 Figure~\ref{fig:results_spectra}. 
 
 We measured the line width of the 
\ion{Fe}{13} 1074 and the \ion{He}{1} 1083 lines by fitting their
spectral shapes after averaging over 3 pixels along the slit \textbf{direction}. 
We fitted the line profiles with Gaussian line shapes and a linear background
to take into account the varying background continuum level.
The results from the line
fitting are shown in panel (e) of Figure~\ref{fig:results_spectra},
where the line width of the \ion{Fe}{13} 1074\,nm and the \ion{He}{1} 
1083\,nm lines are shown as the orange and blue markers, as well as 
the \ion{He}{1} 1083 line intensity shown in green.
The standard deviation of the
inferred parameters, provided from the Levenberg–Marquardt fitting routine, 
are presented as the error bars in the figure. 
We note that on average the coronal 
\ion{Fe}{13} 1074\,nm line width agrees with the previous work by 
\citet{Schad_2023}. We applied Equation 14 from 
\citet{2024ApJ...965...40S} to estimate the plasma temperature responsible 
for the observed line width of the lines, while assuming no thermal Doppler broadening. The
\ion{Fe}{13} 1074 line width corresponds to a coronal formation temperatures of 
$\sim$10$^6$\,K.
However, the observed \ion{He}{1} 1083
line width in the corona (blue dots in Figure~\ref{fig:results_spectra} (e))
is significantly lower than
the line width of the \ion{Fe}{13} 1074\,nm (orange dots in Figure~\ref{fig:results_spectra} panel (f)); based on the aforementioned 
approach, the plasma temperature responsible for
the thermal broadening of the \ion{He}{1} 1083 in the corona is on the 
order of 10$^5$\,K. The intensity of the \ion{He}{1} 1083 is shown as the green dots on Figure~\ref{fig:results_spectra} panel (f). We note that the intensity of the \ion{He}{1} 1083 is 
about two to three times lower than the brightest chromospheric signal in our data, as similarly shown in 
\citet{2019ApJ...877...10J}.
Furthermore, the 
\ion{He}{1} 1083 intensity and line width are increasing on the lunar disc
toward the bottom of the slit, towards the closest location of the flash spectrum on the west limb. This directional increase of the \ion{He}{1} 1083
line intensity corresponds to a decreasing
separation of the observed celestial direction and the penumbral/flash spectrum location.
These evidence, combined with the occurrence of the 
bright \ion{He}{1} 1083 signal contemporaneously with the appearance of the flash spectrum on the west limb, agree with the proposed formation mechanism from \citep{Stellmacher_1974} for a terrestrial atmospheric scattering origin of the observed \ion{He}{1} 1083 signal (and bright chromospheric lines in general), discussed in 
the introduction of this article.

However, the signal detected in our data could arise from a few 
different instrumental issues, that need to be addressed. A possible origin of the bright \ion{He}{1} 1083 could be due to camera bleed or 
scattered light in the spectrograph. The evidence against the camera bleed from the bright prominence hypothesis 
is the lack of \ion{He}{1} 1083 signal over the range of the detector that was not imaging the slit,
seen on Figure~\ref{fig:results_spectra} as the region below $\sim$\,20\,{\arcmin}. We 
left this region of the detector to image the lower part of the slit mount to estimate 
the scattered light in the instrument. The lack of \ion{He}{1} 1083 signal on this
region discredits the possibility of camera sensor bleed issue. 
Furthermore, during the laboratory characterization of the instrument with 
a light source imaged at infinity, significantly brighter than the observed corona, we did not detect any significant light bleed issues. scattered light
if the light source was not aligned with the slit.

Another possible source of the observed \ion{He}{1} 1083 signal is scattering inside of the actual spectrograph. To measure the performance of the spectrograph, we show in Figure~\ref{fig:calibration_data} Panel (b) the spectra from the sky and the on disc calibrations. Note that on the disc observations, the spectral lines have contrast of about 30\% to the continuum, while in the sky spectra, the spectral lines have contrast on the order of 2\% of the continuum. This shows that the scattered light contribution from the disc to the gray scattered light in the instrument is on the order of 15 times less and cannot reproduce the observed \ion{He}{1} 1083 intensity during the eclipse which has almost constant brightness across the sky, which is twice dimmer than the brightest chromospheric signals close to the limb.

\section{Conclusions}
\label{sec:Conclusions}

We designed and built
a near-IR spectrograph with sufficient resolution to observe the 
\ion{He}{1} 1083\,nm line and test the origin of the ``diffuse'' 
neutral helium corona. Testing this hypothesis drove the 
design requirements of CHEESE, in particular the need for sufficient 
spectral resolution, to be able to resolve the thermal width of the line and to observe a significant part 
of the lunar disc and the corona at the same time. We built a classical
slit spectrograph with R\,$\sim$\,9000 in its current setting (depending
on the width of the slit used) as shown in Section~\ref{sec:instrument_design}.

We deployed successfully CHEESE at Dardanelle, AR during the April 8th 2024
total solar eclipse. Our calibration measurements of the solar disc, presented in Figure~\ref{fig:calibration_data}, 
demonstrate that the instrument met its design requirements. 
In particular, we were able to observe the \ion{He}{1} 1083 on the solar disc right
before the eclipse with sufficient spectral resolution to discern it from the 
nearby \ion{Si}{1} 1082.7\,nm line. 
Despite the technical difficulties during the eclipse, described in
Section~\ref{sec:Results}, CHEESE managed to collect data during the eclipse successfully. 
As shown clearly in Figure~\ref{fig:results_spectra}, we observed the coronal 
\ion{Fe}{13} 1074\,nm line, as well as the \ion{He}{1} 1083 and the \ion{H}{1}
Paschen-$\gamma$ 1093.8\,nm. The coronal \ion{Fe}{13} 1074\,nm line
was observed only above the lunar limb in the solar corona,
whereas the strong chromospheric lines were observed both above the 
limb and on the lunar disc. We observed around third contact 
the ubiquitous \ion{He}{1} 1083 across the whole slit, with almost constant intensity
and thermal line width corresponding to chromospheric temperatures.

Similarly to the conclusions in \cite{Stellmacher_1974} for the 
chromospheric lines in the blue part of the visible spectrum,
we believe that we observed during the total solar eclipse
a local (between us and the lunar surface) origin of the 
\ion{He}{1} 1083 (and the 
\ion{H}{1} Paschen-$\gamma$ line) due to their presence on the lunar disc.
The timing of the detection of the \ion{He}{1} 1083 close to third
contact agrees with the previously shown results by \cite{2019ApJ...877...10J}. This evidence, the chromospheric thermal line width of the detected \ion{He}{1} 1083 in the corona and on the lunar disc, and the increase of the \ion{He}{1} 1083 intensity towards the location of 
the approaching flash spectrum support the proposed formation mechanism of terrestrial 
scattering of the flash spectrum radiation as in \cite{Stellmacher_1974}. There are a few other plausible explanations for our observation, which need to be addressed in detail before rejected. We do 
not anticipate an terrestrial exospheric origin of the signal
\citep{lammer_exosphere_2022}, coming 
from \ion{He}{1} ions, because this hypothesis cannot explain the presence
of the previously observed lines of \ion{Ca}{2} ion in the blue part 
of the spectrum~\cite{1955C&T....71..288M}. Furthermore, laboratory 
testing and calibration observations of CHEESE showed that 
scattered light or camera bleed cannot readily explain the observed 
signals, as discussed in Section~\ref{sec:Results}.
However, this observation grants a further repeat of this observations and we believe measurements 
above (much of) the atmosphere, in conditions of
significantly lower sky brightness,
such as with the CORSAIR balloon, would judge a definitive verdict on the
scattering hypothesis we suggest in this work. Constraining the abundance and origin of neutral helium in the heliosphere is crucial for the accurate modeling of the plasma conditions across the heliosphere~\citep{2023ApJ...953..107S}, making the implications from this study applicable beyond the realm of eclipse and atmospheric science.

\begin{acknowledgments}
The authors express their gratitude to the members of the HAO Eclipse Expedition for their support.
MEM would like to especially thank Cory Buhay, Clementine Mitchell, and Vesta 
Alexander-Molnar for their unconditional support with the incessantly arising 
challenges during the expedition. We also want to thank Skylar Shaver
(CU/ LASP) for generously accommodating the HAO eclipse expedition in her home
in Dardanelle, AR. 
The authors would like to thank Dr. Kevin Reardon (NSO), Dr. Alejandro Soto (SwRI), and the anonymous referee for the enriching discussions and suggestions.
The National Center for Atmospheric Research is a
major facility sponsored by the NSF under Cooperative
Agreement No. 1852977. MEM was supported by the ASP Postdoctoral fellowship. 
This work was supported by NSF Grant Award 2504074. 
The K-Cor data is a courtesy of the Mauna Loa Solar Observatory, operated by the
High Altitude Observatory, as part of the National Center for Atmospheric 
Research (NCAR). We would like to thank Dr. Miloslal Dr\"{u}kmuller for his permission to use their white light coronal image taken in Durango, Mexico. The following Python packages were used in this work:
\texttt{astropy}~\citep{astropy_2022};
\texttt{numpy}~\citep{harris2020array};
\texttt{scipy}~\citep{2020SciPy-NMeth};
and \texttt{sunpy}~\citep{sunpy_community2020}. This research has made use of the Astrophysics Data System, funded by NASA under Cooperative Agreement 80NSSC21M00561.

\end{acknowledgments}

%

\vspace{5mm}




\bibliography{CHEESE_Eclipse_2024_results}{}

\begin{thebibliography}{}
\expandafter\ifx\csname natexlab\endcsname\relax\def\natexlab#1{#1}\fi
\providecommand{\url}[1]{\href{#1}{#1}}
\providecommand{\dodoi}[1]{doi:~\href{http://doi.org/#1}{\nolinkurl{#1}}}
\providecommand{\doeprint}[1]{\href{http://ascl.net/#1}{\nolinkurl{http://ascl.net/#1}}}
\providecommand{\doarXiv}[1]{\href{https://arxiv.org/abs/#1}{\nolinkurl{https://arxiv.org/abs/#1}}}

\bibitem[{{Antonucci} {et~al.}(2020){Antonucci}, {Romoli}, {Andretta},
  {Fineschi}, {Heinzel}, {Moses}, {Naletto}, {Nicolini}, {Spadaro}, {Teriaca},
  {Berlicki}, {Capobianco}, {Crescenzio}, {Da Deppo}, {Focardi}, {Frassetto},
  {Heerlein}, {Landini}, {Magli}, {Marco Malvezzi}, {Massone}, {Melich},
  {Nicolosi}, {Noci}, {Pancrazzi}, {Pelizzo}, {Poletto}, {Sasso},
  {Sch{\"u}hle}, {Solanki}, {Strachan}, {Susino}, {Tondello}, {Uslenghi},
  {Woch}, {Abbo}, {Bemporad}, {Casti}, {Dolei}, {Grimani}, {Messerotti},
  {Ricci}, {Straus}, {Telloni}, {Zuppella}, {Auch{\`e}re}, {Bruno},
  {Ciaravella}, {Corso}, {Alvarez Copano}, {Aznar Cuadrado}, {D'Amicis},
  {Enge}, {Gravina}, {Jej{\v{c}}i{\v{c}}}, {Lamy}, {Lanzafame}, {Meierdierks},
  {Papagiannaki}, {Peter}, {Fernandez Rico}, {Giday Sertsu}, {Staub},
  {Tsinganos}, {Velli}, {Ventura}, {Verroi}, {Vial}, {Vives}, {Volpicelli},
  {Werner}, {Zerr}, {Negri}, {Castronuovo}, {Gabrielli}, {Bertacin},
  {Carpentiero}, {Natalucci}, {Marliani}, {Cesa}, {Laget}, {Morea},
  {Pieraccini}, {Radaelli}, {Sandri}, {Sarra}, {Cesare}, {Del Forno}, {Massa},
  {Montabone}, {Mottini}, {Quattropani}, {Schillaci}, {Boccardo}, {Brando},
  {Pandi}, {Baietto}, {Bertone}, {Alvarez-Herrero}, {Garc{\'\i}a Parejo},
  {Cebollero}, {Amoruso}, \& {Centonze}}]{METIS_2020}
{Antonucci}, E., {Romoli}, M., {Andretta}, V., {et~al.} 2020, \aap, 642, A10,
  \dodoi{10.1051/0004-6361/201935338}

\bibitem[{{Astropy Collaboration} {et~al.}(2022){Astropy Collaboration},
  {Price-Whelan}, {Lim}, {Earl}, {Starkman}, {Bradley}, {Shupe}, {Patil},
  {Corrales}, {Brasseur}, {N{\"o}the}, {Donath}, {Tollerud}, {Morris},
  {Ginsburg}, {Vaher}, {Weaver}, {Tocknell}, {Jamieson}, {van Kerkwijk},
  {Robitaille}, {Merry}, {Bachetti}, {G{\"u}nther}, {Aldcroft},
  {Alvarado-Montes}, {Archibald}, {B{\'o}di}, {Bapat}, {Barentsen},
  {Baz{\'a}n}, {Biswas}, {Boquien}, {Burke}, {Cara}, {Cara}, {Conroy},
  {Conseil}, {Craig}, {Cross}, {Cruz}, {D'Eugenio}, {Dencheva}, {Devillepoix},
  {Dietrich}, {Eigenbrot}, {Erben}, {Ferreira}, {Foreman-Mackey}, {Fox},
  {Freij}, {Garg}, {Geda}, {Glattly}, {Gondhalekar}, {Gordon}, {Grant},
  {Greenfield}, {Groener}, {Guest}, {Gurovich}, {Handberg}, {Hart},
  {Hatfield-Dodds}, {Homeier}, {Hosseinzadeh}, {Jenness}, {Jones}, {Joseph},
  {Kalmbach}, {Karamehmetoglu}, {Ka{\l}uszy{\'n}ski}, {Kelley}, {Kern},
  {Kerzendorf}, {Koch}, {Kulumani}, {Lee}, {Ly}, {Ma}, {MacBride}, {Maljaars},
  {Muna}, {Murphy}, {Norman}, {O'Steen}, {Oman}, {Pacifici}, {Pascual},
  {Pascual-Granado}, {Patil}, {Perren}, {Pickering}, {Rastogi}, {Roulston},
  {Ryan}, {Rykoff}, {Sabater}, {Sakurikar}, {Salgado}, {Sanghi}, {Saunders},
  {Savchenko}, {Schwardt}, {Seifert-Eckert}, {Shih}, {Jain}, {Shukla}, {Sick},
  {Simpson}, {Singanamalla}, {Singer}, {Singhal}, {Sinha}, {Sip{\H{o}}cz},
  {Spitler}, {Stansby}, {Streicher}, {{\v{S}}umak}, {Swinbank}, {Taranu},
  {Tewary}, {Tremblay}, {de Val-Borro}, {Van Kooten}, {Vasovi{\'c}}, {Verma},
  {de Miranda Cardoso}, {Williams}, {Wilson}, {Winkel}, {Wood-Vasey}, {Xue},
  {Yoachim}, {Zhang}, {Zonca}, \& {Astropy Project
  Contributors}}]{astropy_2022}
{Astropy Collaboration}, {Price-Whelan}, A.~M., {Lim}, P.~L., {et~al.} 2022,
  \apj, 935, 167, \dodoi{10.3847/1538-4357/ac7c74}

\bibitem[{{Bappu} {et~al.}(1972){Bappu}, {Bhattacharyya}, \&
  {Sivaraman}}]{1972SoPh...26..366B}
{Bappu}, M.~K.~V., {Bhattacharyya}, J.~C., \& {Sivaraman}, K.~R. 1972,
  \solphys, 26, 366, \dodoi{10.1007/BF00165277}

\bibitem[{{Caccin} {et~al.}(1971){Caccin}, {Moschi}, {Rigutti}, \&
  {Falciani}}]{1971SoPh...17...89C}
{Caccin}, B., {Moschi}, G., {Rigutti}, M., \& {Falciani}, R. 1971, \solphys,
  17, 89, \dodoi{10.1007/BF00152863}

\bibitem[{{Casini} {et~al.}(2023){Casini}, {Gibson}, {Bak-Ste{\'s}licka},
  {Newmark}, {Fineschi}, {Fan}, {Auchere}, {Dahlin}, {Casti}, {Karpen},
  {Viall}, {Raouafi}, {Chamberlin}, {Woods}, {Burkepile}, {Farid}, {Gilbert},
  {Cho}, \& {Kim}}]{2023BAAS...55c.047C}
{Casini}, R., {Gibson}, S., {Bak-Ste{\'s}licka}, U., {et~al.} 2023, in Bulletin
  of the American Astronomical Society, Vol.~55, 047,
  \dodoi{10.3847/25c2cfeb.e2c303f5}

\bibitem[{{Caspi} {et~al.}(2023){Caspi}, {Seaton}, {Casini}, {Downs}, {Gibson},
  {Gilbert}, {Glesener}, {Guidoni}, {Hughes}, {McKenzie}, {Plowman}, {Reeves},
  {Saint-Hilaire}, {Shih}, {West}, {Alaoui}, {Alexander}, {Allred}, {Ashfield},
  {Bradshaw}, {Br{\"o}se}, {Buitrago-Casas}, {Chamberlin}, {Chandra}, {Che},
  {Chen}, {Chen}, {Cheng}, {Christe}, {Dahlin}, {de Nolfo}, {Dickson},
  {Dud{\'\i}k}, {Emslie}, {Erdelyi}, {Gallagher}, {Gan}, {Gary}, {Guo},
  {Hayes}, {Hudson}, {Ji}, {Jones}, {Kerr}, {Keshav}, {Knuth}, {Kooi},
  {Kumari}, {Li}, {Limousin}, {Longo}, {Maharana}, {Mandrini}, {Martinez
  Oliveros}, {Massone}, {McAteer}, {McConnell}, {McTiernan}, {Meuris},
  {Mitchell}, {Motorina}, {Mrozek}, {Musset}, {Narukage}, {Nayak}, {Newmark},
  {Nitta}, {Panesar}, {Pariat}, {Piana}, {Qiu}, {Raouafi}, {Reale}, {Rozelot},
  {Samaiyah}, {Sarkar}, {Shaik}, {Skokic}, {Stores}, {Struminsky}, {Su},
  {Takahashi}, {Tiwari}, {Tsiklauri}, {Vievering}, {Warmuth}, {White}, {Zhang},
  \& {Zimovets}}]{2023BAAS...55c.048C}
{Caspi}, A., {Seaton}, D., {Casini}, R., {et~al.} 2023, in Bulletin of the
  American Astronomical Society, Vol.~55, 048,
  \dodoi{10.3847/25c2cfeb.b95dd671}

\bibitem[{{Centeno} {et~al.}(2008){Centeno}, {Trujillo Bueno}, {Uitenbroek}, \&
  {Collados}}]{Centeno_2008}
{Centeno}, R., {Trujillo Bueno}, J., {Uitenbroek}, H., \& {Collados}, M. 2008,
  \apj, 677, 742, \dodoi{10.1086/528680}

\bibitem[{{Colacevich}(1952)}]{Colacevic_1956}
{Colacevich}, A. 1952, {\it Academia Nazionale dei Lincei, Fondazione Volta},
  186

\bibitem[{de~Wijn {et~al.}(2012)de~Wijn, Burkepile, Tomczyk, Nelson, Huang, \&
  Gallagher}]{deWijn_2012}
de~Wijn, A.~G., Burkepile, J.~T., Tomczyk, S., {et~al.} 2012, in Ground-based
  and Airborne Telescopes IV, ed. L.~M. Stepp, R.~Gilmozzi, \& H.~J. Hall, Vol.
  8444, International Society for Optics and Photonics (SPIE), 84443N,
  \dodoi{10.1117/12.926511}

\bibitem[{{Del Zanna} {et~al.}(2020){Del Zanna}, {Storey}, {Badnell}, \&
  {Andretta}}]{2020ApJ...898...72D}
{Del Zanna}, G., {Storey}, P.~J., {Badnell}, N.~R., \& {Andretta}, V. 2020,
  \apj, 898, 72, \dodoi{10.3847/1538-4357/ab9d84}

\bibitem[{{Dellbouille} {et~al.}(1981){Dellbouille}, {Roland}, {Brault}, \&
  {Testerman}}]{Dellbouille_IR_solar_atlas}
{Dellbouille}, L., {Roland}, G., {Brault}, J., \& {Testerman}, L. 1981,
  {Photometric Atlas of the Solar Spectrum from 1.850 to 10.000 cm$^{-1}$}
  ({Tucson, Arizona, USA})

\bibitem[{{Deutsch} \& {Righini}(1964)}]{1964ApJ...140..313D}
{Deutsch}, A.~J., \& {Righini}, G. 1964, \apj, 140, 313, \dodoi{10.1086/147920}

\bibitem[{{Dima} {et~al.}(2016){Dima}, {Kuhn}, \& {Berdyugina}}]{Dima_2016}
{Dima}, G., {Kuhn}, J., \& {Berdyugina}, S. 2016, {\it Frontiers in Astronomy
  and Space Sciences}, 3, 13, \dodoi{10.3389/fspas.2016.00013}

\bibitem[{{Ding} \& {Habbal}(2017)}]{2017ApJ...842L...7D}
{Ding}, A., \& {Habbal}, S.~R. 2017, \apjl, 842, L7,
  \dodoi{10.3847/2041-8213/aa7460}

\bibitem[{{Fan}(2018)}]{2018ApJ...862...54F}
{Fan}, Y. 2018, \apj, 862, 54, \dodoi{10.3847/1538-4357/aaccee}

\bibitem[{{Grotrian}(1934)}]{Grotrian_1934}
{Grotrian}, W. 1934, {\it Zeitschrift fuer Astrophysik}, 8, 124

\bibitem[{Harris {et~al.}(2020)Harris, Millman, van~der Walt, Gommers,
  Virtanen, Cournapeau, Wieser, Taylor, Berg, Smith, Kern, Picus, Hoyer, van
  Kerkwijk, Brett, Haldane, del R{\'{i}}o, Wiebe, Peterson,
  G{\'{e}}rard-Marchant, Sheppard, Reddy, Weckesser, Abbasi, Gohlke, \&
  Oliphant}]{harris2020array}
Harris, C.~R., Millman, K.~J., van~der Walt, S.~J., {et~al.} 2020, \nat, 585,
  357, \dodoi{10.1038/s41586-020-2649-2}

\bibitem[{{Heinzel} {et~al.}(2023){Heinzel}, {Jej{\v{c}}i{\v{c}}},
  {{\v{S}}t{\v{e}}p{\'a}n}, {Susino}, {Andretta}, {Russano}, {Fineschi},
  {Romoli}, {Bemporad}, {Berlicki}, {Burtovoi}, {Da Deppo}, {De Leo},
  {Grimani}, {Jerse}, {Landini}, {Naletto}, {Nicolini}, {Pancrazzi}, {del Pino
  Alem{\'a}n}, {Sasso}, {Spadaro}, {Stangalini}, {Telloni}, {Teriaca},
  {Uslenghi}, \& {Ar{\'e}valo}}]{2023ApJ...957L..10H}
{Heinzel}, P., {Jej{\v{c}}i{\v{c}}}, S., {{\v{S}}t{\v{e}}p{\'a}n}, J., {et~al.}
  2023, \apjl, 957, L10, \dodoi{10.3847/2041-8213/acff62}

\bibitem[{{Judge} {et~al.}(2019){Judge}, {Tomczyk}, {Hannigan}, \&
  {Sewell}}]{2019ApJ...877...10J}
{Judge}, P., {Tomczyk}, S., {Hannigan}, J., \& {Sewell}, S. 2019, \apj, 877,
  10, \dodoi{10.3847/1538-4357/ab0e04}

\bibitem[{{Khan} {et~al.}(2024){Khan}, {Gibson}, {Casini}, \&
  {Nagaraju}}]{2024ApJ...971...27K}
{Khan}, R., {Gibson}, S.~E., {Casini}, R., \& {Nagaraju}, K. 2024, \apj, 971,
  27, \dodoi{10.3847/1538-4357/ad55ed}

\bibitem[{{Khan} \& {Nagaraju}(2022)}]{2022SoPh..297...96K}
{Khan}, R., \& {Nagaraju}, K. 2022, \solphys, 297, 96,
  \dodoi{10.1007/s11207-022-02024-2}

\bibitem[{{Koutchmy} {et~al.}(2019){Koutchmy}, {Baudin}, {Abdi}, {Golub}, \&
  {S{\`e}vre}}]{2019A&A...632A..86K}
{Koutchmy}, S., {Baudin}, F., {Abdi}, S., {Golub}, L., \& {S{\`e}vre}, F. 2019,
  \aap, 632, A86, \dodoi{10.1051/0004-6361/201935681}

\bibitem[{{Kuhn} {et~al.}(2007){Kuhn}, {Arnaud}, {Jaeggli}, {Lin}, \&
  {Moise}}]{2007ApJ...667L.203K}
{Kuhn}, J.~R., {Arnaud}, J., {Jaeggli}, S., {Lin}, H., \& {Moise}, E. 2007,
  \apjl, 667, L203, \dodoi{10.1086/522370}

\bibitem[{{Kuhn} {et~al.}(1996){Kuhn}, {Penn}, \& {Mann}}]{1996ApJ...456L..67K}
{Kuhn}, J.~R., {Penn}, M.~J., \& {Mann}, I. 1996, \apjl, 456, L67,
  \dodoi{10.1086/309864}

\bibitem[{Lammer {et~al.}(2022)Lammer, Scherf, Ito, Mura, Vorburger, Guenther,
  Wurz, Erkaev, \& Odert}]{lammer_exosphere_2022}
Lammer, H., Scherf, M., Ito, Y., {et~al.} 2022, \ssr, 218, 15,
  \dodoi{10.1007/s11214-022-00876-5}

\bibitem[{{Landi Degl'Innocenti}(1982)}]{1982SoPh...79..291L}
{Landi Degl'Innocenti}, E. 1982, \solphys, 79, 291, \dodoi{10.1007/BF00146246}

\bibitem[{{Lemen} {et~al.}(2012){Lemen}, {Title}, {Akin}, {Boerner}, {Chou},
  {Drake}, {Duncan}, {Edwards}, {Friedlaender}, {Heyman}, {Hurlburt}, {Katz},
  {Kushner}, {Levay}, {Lindgren}, {Mathur}, {McFeaters}, {Mitchell}, {Rehse},
  {Schrijver}, {Springer}, {Stern}, {Tarbell}, {Wuelser}, {Wolfson}, {Yanari},
  {Bookbinder}, {Cheimets}, {Caldwell}, {Deluca}, {Gates}, {Golub}, {Park},
  {Podgorski}, {Bush}, {Scherrer}, {Gummin}, {Smith}, {Auker}, {Jerram},
  {Pool}, {Soufli}, {Windt}, {Beardsley}, {Clapp}, {Lang}, \&
  {Waltham}}]{2012SoPh..275...17L}
{Lemen}, J.~R., {Title}, A.~M., {Akin}, D.~J., {et~al.} 2012, \solphys, 275,
  17, \dodoi{10.1007/s11207-011-9776-8}

\bibitem[{{Migeotte} \& {Rosen}(1955)}]{1955C&T....71..288M}
{Migeotte}, M., \& {Rosen}, B. 1955, {\it Ciel et Terre}, 71, 288

\bibitem[{{Moise} {et~al.}(2010){Moise}, {Raymond}, \&
  {Kuhn}}]{2010ApJ...722.1411M}
{Moise}, E., {Raymond}, J., \& {Kuhn}, J.~R. 2010, \apj, 722, 1411,
  \dodoi{10.1088/0004-637X/722/2/1411}

\bibitem[{{Molnar} \& {Casini}(2024)}]{Molnar_2024}
{Molnar}, M.~E., \& {Casini}, R. 2024, \apj, 977, 97,
  \dodoi{10.3847/1538-4357/ad8de4}

\bibitem[{{M{\"u}ller} {et~al.}(2020){M{\"u}ller}, {St. Cyr}, {Zouganelis},
  {Gilbert}, {Marsden}, {Nieves-Chinchilla}, {Antonucci}, {Auch{\`e}re},
  {Berghmans}, {Horbury}, {Howard}, {Krucker}, {Maksimovic}, {Owen}, {Rochus},
  {Rodriguez-Pacheco}, {Romoli}, {Solanki}, {Bruno}, {Carlsson}, {Fludra},
  {Harra}, {Hassler}, {Livi}, {Louarn}, {Peter}, {Sch{\"u}hle}, {Teriaca}, {del
  Toro Iniesta}, {Wimmer-Schweingruber}, {Marsch}, {Velli}, {De Groof},
  {Walsh}, \& {Williams}}]{SolO_2020}
{M{\"u}ller}, D., {St. Cyr}, O.~C., {Zouganelis}, I., {et~al.} 2020, \aap, 642,
  A1, \dodoi{10.1051/0004-6361/202038467}

\bibitem[{{Muro} {et~al.}(2023){Muro}, {Gunn}, {Fearn}, {Fearn}, \&
  {Morgan}}]{2023SoPh..298...75M}
{Muro}, G.~D., {Gunn}, M., {Fearn}, S., {Fearn}, T., \& {Morgan}, H. 2023,
  \solphys, 298, 75, \dodoi{10.1007/s11207-023-02162-1}

\bibitem[{{Neckel}(1999)}]{1999SoPh..184..421N}
{Neckel}, H. 1999, \solphys, 184, 421, \dodoi{10.1023/A:1017165208013}

\bibitem[{{Raouafi} {et~al.}(1999){Raouafi}, {Lemaire}, \&
  {Sahal-Br{\'e}chot}}]{1999A&A...345..999R}
{Raouafi}, N.~E., {Lemaire}, P., \& {Sahal-Br{\'e}chot}, S. 1999, \aap, 345,
  999

\bibitem[{{Raouafi} {et~al.}(2016){Raouafi}, {Riley}, {Gibson}, {Fineschi}, \&
  {Solanki}}]{2016FrASS...3...20R}
{Raouafi}, N.~E., {Riley}, P., {Gibson}, S., {Fineschi}, S., \& {Solanki},
  S.~K. 2016, {\it Frontiers in Astronomy and Space Sciences}, 3, 20,
  \dodoi{10.3389/fspas.2016.00020}

\bibitem[{{Raouafi} {et~al.}(2002){Raouafi}, {Sahal-Br{\'e}chot}, \&
  {Lemaire}}]{2002A&A...396.1019R}
{Raouafi}, N.~E., {Sahal-Br{\'e}chot}, S., \& {Lemaire}, P. 2002, \aap, 396,
  1019, \dodoi{10.1051/0004-6361:20021418}

\bibitem[{{Schad} {et~al.}(2023){Schad}, {Kuhn}, {Fehlmann}, {Scholl},
  {Harrington}, {Rimmele}, \& {Tritschler}}]{Schad_2023}
{Schad}, T.~A., {Kuhn}, J.~R., {Fehlmann}, A., {et~al.} 2023, \apj, 943, 59,
  \dodoi{10.3847/1538-4357/acabbd}

\bibitem[{{Schad} {et~al.}(2024){Schad}, {Fehlmann}, {Dima}, {Kuhn}, {Scholl},
  {Harrington}, {Rimmele}, {Tritschler}, \& {Paraschiv}}]{2024ApJ...965...40S}
{Schad}, T.~A., {Fehlmann}, A., {Dima}, G.~I., {et~al.} 2024, \apj, 965, 40,
  \dodoi{10.3847/1538-4357/ad2995}

\bibitem[{{Stellmacher} \& {Koutchmy}(1974)}]{Stellmacher_1974}
{Stellmacher}, G., \& {Koutchmy}, S. 1974, \aap, 35, 43

\bibitem[{{Swaczyna} {et~al.}(2023){Swaczyna}, {Bzowski}, {Heerikhuisen},
  {Kubiak}, {Rahmanifard}, {Zirnstein}, {Fuselier}, {Galli}, {McComas},
  {M{\"o}bius}, \& {Schwadron}}]{2023ApJ...953..107S}
{Swaczyna}, P., {Bzowski}, M., {Heerikhuisen}, J., {et~al.} 2023, \apj, 953,
  107, \dodoi{10.3847/1538-4357/ace719}

\bibitem[{{The SunPy Community} {et~al.}(2020){The SunPy Community}, Barnes,
  Bobra, Christe, Freij, Hayes, Ireland, Mumford, Perez-Suarez, Ryan, Shih,
  Chanda, Glogowski, Hewett, Hughitt, Hill, Hiware, Inglis, Kirk, Konge, Mason,
  Maloney, Murray, Panda, Park, Pereira, Reardon, Savage, Sipőcz, Stansby,
  Jain, Taylor, Yadav, Rajul, \& Dang}]{sunpy_community2020}
{The SunPy Community}, Barnes, W.~T., Bobra, M.~G., {et~al.} 2020, \apj, 890,
  68, \dodoi{10.3847/1538-4357/ab4f7a}

\bibitem[{{Tomczyk} {et~al.}(2008){Tomczyk}, {Card}, {Darnell}, {Elmore},
  {Lull}, {Nelson}, {Streander}, {Burkepile}, {Casini}, \&
  {Judge}}]{2008SoPh..247..411T}
{Tomczyk}, S., {Card}, G.~L., {Darnell}, T., {et~al.} 2008, \solphys, 247, 411,
  \dodoi{10.1007/s11207-007-9103-6}

\bibitem[{Virtanen {et~al.}(2020)Virtanen, Gommers, Oliphant, Haberland, Reddy,
  Cournapeau, Burovski, Peterson, Weckesser, Bright, {van der Walt}, Brett,
  Wilson, Millman, Mayorov, Nelson, Jones, Kern, Larson, Carey, Polat, Feng,
  Moore, {VanderPlas}, Laxalde, Perktold, Cimrman, Henriksen, Quintero, Harris,
  Archibald, Ribeiro, Pedregosa, {van Mulbregt}, \& {SciPy 1.0
  Contributors}}]{2020SciPy-NMeth}
Virtanen, P., Gommers, R., Oliphant, T.~E., {et~al.} 2020, {\it Nature
  Methods}, 17, 261, \dodoi{10.1038/s41592-019-0686-2}

\bibitem[{{Yang} {et~al.}(2020){Yang}, {Bethge}, {Tian}, {Tomczyk}, {Morton},
  {Del Zanna}, {McIntosh}, {Karak}, {Gibson}, {Samanta}, {He}, {Chen}, \&
  {Wang}}]{2020Sci...369..694Y}
{Yang}, Z., {Bethge}, C., {Tian}, H., {et~al.} 2020, \sci, 369, 694,
  \dodoi{10.1126/science.abb4462}

\bibitem[{{Zhu} {et~al.}(2024){Zhu}, {Habbal}, {Ding}, {Yamashiro}, {Landi},
  {Boe}, {Constantinou}, \& {Nassir}}]{2024ApJ...966..122Z}
{Zhu}, Y., {Habbal}, S.~R., {Ding}, A., {et~al.} 2024, \apj, 966, 122,
  \dodoi{10.3847/1538-4357/ad3424}

\end{thebibliography}
\bibliographystyle{aasjournal}



\end{document}